\documentclass{article}
\usepackage{a4,amsmath}

\begin{document}

\title{AN ALTERNATIVE MODEL FOR THE DUFFIN-KEMMER-PETIAU OSCILLATOR}

\author{ D. A. Kulikov, R. S. Tutik\thanks{tutik@ff.dsu.dp.ua}
\quad and A. P. Yaroshenko,\\
\\
Dniepropetrovsk National University, \\
Kazakova Str. 20, Dniepropetrovsk, 49050,
Ukraine\\
}

\maketitle

\begin{abstract}

\large
A new oscillator model with different form of the non-minimal substitution
within the framework of the Duffin-Kemmer-Petiau equation is offered.
The model possesses exact solutions and a discrete spectrum of high degeneracy.
The distinctive property of the proposed model is the lack of the spin-orbit
interaction, being typical for other relativistic models with the non-minimal
substitution, and the different value of the zero-point energy in comparison
with that for the Duffin-Kemmer-Petiau oscillator described in the literature.

\vspace{5mm}

\vspace{5mm}
\end{abstract}

\newpage

\normalsize
\section{Introduction}

The harmonic oscillator is one of the most basic and useful solvable examples
in the non-relativistic quantum mechanics. Recently, the harmonic-oscillator potential
for relativistic spin-1/2 particles has received considerable attention by many groups.
Besides the academic importance of the exact solutions, the relativistic oscillator may be
an analytical basis to deal with more realistic interactions in several areas, namely,
in nuclear and particle physics. In addition to the exact solutions of the Dirac equation
with confining scalar-vector potentials \cite{b1,b2,b3,b4}, since the paper by Moshinsky
and Szczepaniak  \cite{b5}, the interest has been extended to the oscillator model constructed
by means of a non-minimal coupling scheme.

This model, introduced formerly in the works  \cite{b6,b7,b8,b9},
has the Hamiltonian which is linear in both momenta
and coordinates, conserves angular momentum, is exactly solvable and
its eigenspectrum is highly degenerate. The system obtained, being a harmonic
oscillator with a spin-orbit coupling in the non-relativistic limit, is referred
to as the Dirac oscillator (for further references see Ref. 10).

Analogous models have been constructed within the framework of the Klein-Gordon
equation  \cite{b11,b12}, and the spin-0 and spin-1 Duffin-Kemmer-Petiau (DKP) formalism,
too. In the latter case, there are two different approaches. The first approach  \cite{b13,b14},
the PSUSY oscillator, employs ideas of the parasupersymmetric quantum mechanics \cite{b15}
whereas the second one  \cite{b16}, the DKP oscillator, exploits the technique developed
for the Dirac oscillator  \cite{b5}. However, the PSUSY oscillator has been proved
to be equivalent to
the DKP oscillator supplemented with the additional constraint  \cite{b17}.

The eigenstates of the spin-1 DKP oscillator subdivide into the states with the
natural and unnatural parities, whereas for the PSUSY oscillator only the
natural-parity states become permissible due to the above-mentioned constraint.
It should be pointed out that the energy eigenvalues of unnatural-parity states of
the spin-1 DKP oscillator show the behavior which is unusual for oscillator models.
These energy eigenvalues are not equidistant in contrast to those of the spin-0
and spin-1/2 oscillators and the spectrum of the spin-1 oscillator with the natural-parity
states.

The goal of the present work is to study a more general type of the non-minimal
substitution allowed within the framework of the DKP equation and to show that
there is an alternative
model of the spin-1 DKP oscillator whose energy spectrum is equidistant for the states
of both the natural and unnatural parity.

\section{Generalization of the Model of the Duffin-Kemmer-Petiau Oscillator}

In this section we sketch out the technique of the non-minimal
substitution and consider the generalization of the model of the DKP oscillator.

Let us recall that the Duffin-Kemmer-Petiau equation \cite{b18,b19,b20}
describing a free scalar or vector boson of the nonzero mass $m$ can
be written as
\begin{equation}\label{eq1}
i \hbar\beta^{0}\frac{\partial\psi}{\partial t}=(c\vec{\beta}\cdot\vec{p}
+mc^{2})\psi,
\end{equation}
where the matrices $\beta^{\mu}$ ($\mu=0,1,2,3$) satisfy the Kemmer trilinear algebra
\begin{equation}\label{eq2}
\beta^{\mu}\beta^{\nu}\beta^{\lambda}+\beta^{\lambda}\beta^{\nu}\beta^{\mu}=
g^{\mu\nu}\beta^{\lambda}+g^{\lambda\nu}\beta^{\mu}
\end{equation}
and the convention for the metric tensor is
$g^{\mu\nu}=diag(1,-1,-1,-1)$.

Actually, there exist different representations for this algebra \cite{b21}.
However, the most used among them are the following three very well known cases:
the trivial one with $\beta^{\mu}=0$, a five-dimensional representation describing the spin-0
particles, and a ten-dimensional one for the spin-1 bosons. In the last
case, the matrices $\beta^{\mu}$ in the notation of the paper \cite{b16} are given by
\begin{equation}\label{eq3}
 \beta^{0}=\left(\begin{array}{cccc}
   0 & \bar 0 & \bar 0  & \bar 0\\
  \bar 0^{T} & \mathbf{0} & \mathbf{I}  &\mathbf{0}\\
  \bar 0^{T} & \mathbf{I} & \mathbf{0}  & \mathbf{0}\\
  \bar 0^{T} & \mathbf{0} & \mathbf{0}  & \mathbf{0}
  \end{array}\right),
\quad
\beta^{i}=\left(\begin{array}{cccc}
  0 & \bar 0 & e_i  & \bar 0\\
  \bar 0^{T} & \mathbf{0} & \mathbf{0}  & -i s_i\\
  -e_{i}^{T} & \mathbf{0} & \mathbf{0}  & \mathbf{0}\\
  \bar 0^{T} & -i s_i & \mathbf{0}  & \mathbf{0}
  \end{array}\right),
\quad i=1,2,3
\end{equation}
where $s_i$ are the usual $3 \times 3$ spin-1 matrices and
\begin{equation}\label{eq4}
 \bar 0=(\begin{array}{ccc}
  0 &0 & 0
\end{array})
\quad
e_1=(\begin{array}{ccc}
  1 & 0 & 0
\end{array})
\quad
e_2=(\begin{array}{ccc}
  0 & 1 & 0
\end{array})
\quad
e_3=(\begin{array}{ccc}
  0 & 0 & 1
\end{array})
\end{equation}
while $\mathbf{0}$ and $\mathbf{I}$ stand for the $3 \times 3$ zero and identity
matrices, respectively.

Inserting the non-minimal substitution
\begin{equation}\label{eq5}
\vec{p} \rightarrow \vec{p}-i m\omega\eta_{0}\vec{r}
\end{equation}
into (\ref{eq1}) we obtain the equation for the DKP oscillator \cite{b16}
\begin{equation}\label{eq6}
i \hbar\beta^{0}\frac{\partial\psi}{\partial t}=[c\vec{\beta}\cdot
(\vec{p}-i m\omega\eta_{0}\vec{r})+mc^{2}]\psi
\end{equation}
where $\eta_{0}=2(\beta^{0})^{2}-1$ and $\omega$ is the oscillator
frequency.

We attempt to generalize this model through a
modification of the non-minimal substitution. For this purpose we
extend the coupling term to the more general form
\begin{equation}\label{eq7}
\vec{p}\rightarrow \vec{p}-i m\omega M\vec{r},
\end{equation}
where $M$ is some $10 \times 10$ matrix consisted of combinations of
matrices $\beta^{\mu}$.

Obviously, not every matrix $M$ will lead us to an oscillator model.
As it was noticed in Ref. 12, in models of both the spin-0 and
spin-1/2 oscillators, a matrix involving a potential in the
non-minimal substitution anticommutes with the matrix structure of the
momentum part of the wave equation. We suppose that this property
remains valid as well for a model of the spin-1 oscillator. In our
case, the such matrix is $M$, so that the following condition must be
held
\begin{equation}\label{eq8}
\vec{\beta} M+M\vec{\beta}=0.
\end{equation}

For the Kemmer algebra (\ref{eq2}), there exist four matrices anticommuting
with $\vec{\beta}$. Hence, $M$ may be chosen as their linear combination
\begin{equation}\label{eq9}
M=a_{1}\eta_{0}+a_{2}\eta_{5}+a_{3}\eta+a_{4}\eta_{0}\eta_{5}\eta,
\end{equation}
where $a_i$ are numerical coefficients and the matrices $\eta_{5}$ and $\eta$ are given by
\begin{eqnarray}\label{eq10}
\eta_{5}&=&-(2(\beta^{0})^{2}-1)(2(\beta^{1})^{2}+1)(2(\beta^{2})^{2}+1)(2(\beta^{3})^{2}+1),\nonumber\\
\eta&=&2\beta^{\mu}\beta_{\mu}-5.
\end{eqnarray}
Exploiting the spin-1 representation for $\beta^{\mu}$ we have
\begin{equation}\label{eq11}
\eta=-\eta_{5}, \quad \eta_{0}\eta_{5}\eta=-\eta_{0},
\end{equation}
and the final expression for the matrix $M$ reduces to
\begin{equation}\label{eq12}
M=b_{1}\eta_{0}+b_{2}\eta_{5}.
\end{equation}

Thus, the DKP equation with the non-minimal substitution (\ref{eq7})
which includes the matrix $M$ in the form (\ref{eq12}) is written as
\begin{equation}\label{eq13}
i \hbar\beta^{0}\frac{\partial\psi}{\partial t}=[c\vec{\beta}\cdot
(\vec{p}-i m\omega(b_{1}\eta_{0}+b_{2}\eta_{5})\vec{r})+mc^{2}]\psi.
\end{equation}

Next, the block structure (\ref{eq3}) of the matrices  $\beta^{\mu}$
implies that the ten-component wave function of a stationary state
may be decomposed into
\begin{equation}\label{eq14}
\psi (\vec{r}, t) =(\begin{array}{cccc}
  i\varphi(\vec{r}) & \vec{F}(\vec{r}) & \vec{G}(\vec{r}) & \vec{K}(\vec{r})
\end{array})^{T} \exp{(-i Et/\hbar)}
\end{equation}
and the equation (\ref{eq13}) gives rise to the following set of
time-independent equations
\begin{eqnarray} \label{eq15}
mc^{2}\varphi &=& i c(\vec{p}-i(b_{1}+b_{2})m\omega\vec{r})\cdot\vec{G},\nonumber\\
mc^{2}\vec{F} &=& E\vec{G}-c(\vec{p}+i(b_{1}-b_{2})m\omega\vec{r})\times\vec{K},\nonumber\\
mc^{2}\vec{G} &=& E\vec{F}+i c(\vec{p}+i(b_{1}+b_{2})m\omega\vec{r})\varphi,\nonumber\\
mc^{2}\vec{K} &=&-c(\vec{p}-i(b_{1}-b_{2})m\omega\vec{r})\times\vec{F}.
\end{eqnarray}
Upon eliminating $\varphi(\vec{r})$, $\vec{G}(\vec{r})$ and $\vec{K}(\vec{r})$
this set becomes equivalent to the only one equation for the
3-component spinor $\vec{F}(\vec{r})$
\begin{eqnarray}\label{eq16}
 &&(E^{2}-m^{2}c^{4})\vec{F}=[c^{2}(\vec{p}^{2}+(b_{1}-b_{2})^{2}m^{2}\omega^{2}\vec{r}^{2})
\nonumber\\
&&-(3b_{1}-b_{2})\hbar\omega mc^{2}
-2b_{1}\omega mc^{2}\vec{L}\cdot\vec{s}]\vec{F}
+4b_{1}b_{2}c^{2}m^{2}\omega^{2}\vec{r}(\vec{r}\cdot\vec{F})\nonumber\\
&&-\frac{2b_{1}\omega}{m}(\vec{p}+i(b_{1}+b_{2})m\omega\vec{r})
[(\vec{p}-i(b_{1}-b_{2})m\omega\vec{r})\cdot
(2\hbar+\vec{L}\cdot\vec{s})\vec{F}],
\end{eqnarray}
where $\vec{L}$ is the orbital angular momentum and $\vec{s}$ is the $3 \times
3$ spin-1 operator.

The derived equation, apart from the last term, is that
for a harmonic oscillator. However, as $c\rightarrow \infty$ this
last
term becomes negligible and in the non-relativistic limit the described system
corresponds to the ordinary harmonic oscillator of frequency $\omega$
with a strong spin-orbit coupling term. This justifies the name of
the Duffin-Kemmer-Petiau oscillator for our model as well.

\section{Exact solutions to the derived equation}

Because within the framework of the DKP equation
the total angular momentum $\vec{J}=\vec{L}+\hbar\vec{s}$
is conserved,
the spatial variables for the components of the wave function
(\ref{eq14}) are separated and we can write \cite{b22}
\begin{eqnarray}\label{eq17}
\varphi=\frac{1}{r}\phi(r)Y_{JM}(\Omega), \qquad\qquad\quad
\vec{F}=\frac{1}{r}\sum_{L} f_{L}(r)\vec{Y}^{M}_{JL1}(\Omega),\nonumber\\
\vec{G}=\frac{1}{r}\sum_{L} g_{L}(r)\vec{Y}^{M}_{JL1}(\Omega), \qquad
\vec{K}=\frac{1}{r}\sum_{L} k_{L}(r)\vec{Y}^{M}_{JL1}(\Omega),
\end{eqnarray}
where $\vec{Y}^{M}_{JL1}(\Omega)$ are the so-called vector spherical harmonics.

Substituting (\ref{eq17}) into (\ref{eq15}) leads to
ten radial equations which reduce to the two uncoupled sets associated with the
$(-1)^{J}$ and $(-1)^{J+1}$ parities. The $(-1)^{J}$ solutions correspond
to the natural-parity states. They are described by the functions (\ref{eq17})
with the nonzero components $f_{J},g_{J},k_{J\pm 1}$. The orbital angular momentum
for these states has a definite value $J=L$. The $(-1)^{J+1}$ solutions, with values
of $J=L-1$ and $J=L+1$ being mixed, are referred to as the unnatural-parity states.
Their nonzero components are $f_{J\pm 1},g_{J\pm 1},k_{J},\phi$.

In the case of the natural-parity states, the radial equation has
the oscillator-like form
\begin{eqnarray}\label{eq18}
&&( \hbar^{2}\frac{\mathrm{d}^{2}}{\mathrm{d} r^{2}}+\frac{E^{2}-m^{2}c^{4}}{c^{2}}
+(b_{1}-b_{2})\hbar\omega m
\nonumber\\
&&-(b_{1}+b_{2})^{2}m^{2}\omega^{2}r^{2}
-\frac{\hbar^{2} J(J+1)}{r^{2}} ) f_{J}(r)=0,
\end{eqnarray}
whereas for the unnatural-parity states one gets a system of two coupled equations
\begin{eqnarray}\label{eq19}
(\hbar^{2}\frac{\mathrm{d}^{2}}{\mathrm{d} r^{2}}&+&\frac{E^{2}-m^{2}c^{4}}{c^{2}}
-3(b_{1}+b_{2})\hbar\omega m-(b_{1}+b_{2})^{2}m^{2}\omega^{2}r^{2}
-\frac{\hbar^{2} J(J+1)}{r^{2}})\phi(r)\nonumber\\
&-&\frac{2b_{1}\sqrt{J(J+1)}\hbar\omega E}{c^{2}}k_{J}(r)=0,\nonumber\\
(\hbar^{2}\frac{\mathrm{d}^{2}}{\mathrm{d} r^{2}}&+&\frac{E^{2}-m^{2}c^{4}}{c^{2}}
-(b_{1}-b_{2})\hbar\omega m-(b_{1}-b_{2})^{2}m^{2}\omega^{2}r^{2}
-\frac{\hbar^{2} J(J+1)}{r^{2}})k_{J}(r)\nonumber\\
&-&\frac{2b_{1}\sqrt{J(J+1)}\hbar\omega E}{c^{2}}\phi(r)=0.
\end{eqnarray}

Now we consider exact solutions to the last equations. It is easily seen that the
obtained system decouples into two oscillator-like equations  and,
hence, has exact solutions if and only if $b_{1}=0$ or $b_{2}=0$.
For these two cases we have:

(i) $b_{1}=1, \quad b_{2}=0$ (the value $b_{1}=const\neq 1$ can be absorbed
into $\omega$ providing  $b_{1}=1$). This choice of coefficients
reproduces the DKP oscillator model proposed in Ref. 16.
According to (\ref{eq18}) and (\ref{eq19}), its eigenvalues take the form
\begin{eqnarray}\label{eq20}
 E^{2}-m^{2}c^{4}&=&(4n+2J+2)\hbar\omega mc^{2}   (L=J),  \nonumber\\
 E^{2}-m^{2}c^{4}+\hbar\omega\sqrt{m^{2}c^{4}+4J(J+1)E^{2}}&=&(4n+2J+3)\hbar\omega mc^{2} (L=J\pm 1),\nonumber\\
 E^{2}-m^{2}c^{4}-\hbar\omega\sqrt{m^{2}c^{4}+4J(J+1)E^{2}}&=&(4n+2J+3)\hbar\omega mc^{2}
 (L=J\pm 1),
\end{eqnarray}
where $n$ is the radial quantum number. Here the first equation
describes the energies of the natural-parity states and the
rest equations correspond to the
unnatural-parity states.

(ii) $b_{1}=0, \quad b_{2}=1$. With these coefficients we arrive at
the alternative model for the DKP oscillator. Its eigenvalues are
given by
\begin{eqnarray}\label{eq21}
E^{2}-m^{2}c^{4}&=&(4n+2J+4)\hbar\omega mc^{2}  (L=J),\nonumber\\
E^{2}-m^{2}c^{4}&=&(4n+2J+6)\hbar\omega mc^{2}  (L=J\pm 1),\nonumber\\
E^{2}-m^{2}c^{4}&=&(4n+2J+2)\hbar\omega mc^{2}  (L=J\pm 1)
\end{eqnarray}
that can be rewritten with the common formula
\begin{equation}\label{eq22}
E^{2}-m^{2}c^{4}=(4n+2L+4)\hbar\omega mc^{2}\quad (L=J-1,J,J+1).
\end{equation}

In contrast to the equations (\ref{eq20}), the eigenspectrum of this
model proved to be completely equidistant. Besides, these models
differ in the zero-point energy.

Note, that in the case $J=0$, the radial equations should be deduced individually
with putting $f_{-1}(r)=g_{-1}(r)=k_{-1}(r)=0$ into (\ref{eq17}).
For all this the expression (\ref{eq22}) for the eigenvalues
with $L=J+1=1$ remains still valid.

\section{Discussion}

Thus, we have considered the DKP equation with the interaction
introduced through the non-minimal substitution of general form. This
allows us to find a new exactly solvable model for the spin-1 oscillator that
possesses the completely equidistant eigenspectrum.

Moreover, the eigenvalues of the proposed model differ from those of
the spin-0 oscillator \cite{b11,b12} in appearance only the constant
addition term $4\hbar\omega$. It means that within the framework of
the alternative model the spin-1 particle behaves like the spinless
one. In fact, as soon as we insert $b_{1}=0$ into (\ref{eq16}) the
terms with the spin-orbit coupling vanish there.

Hence, it appears that the exactly solvable models for the DKP
oscillator considered in the previous section differ dramatically
from each other. This should be regarded as arising from the
different Lorentz structure of the oscillator coupling terms in these
models.

For elucidating this difference let us turn to the Lorentz covariant
form of the DKP equation (\ref{eq13})
\begin{equation}\label{eq23}
[c\beta^{\mu}p_{\mu}+\frac{1}{2}b_{1}(S_{\mu\nu}\beta_{\lambda}+
\beta_{\lambda}S_{\mu\nu})T^{\mu\nu\lambda}-i b_{2}
\beta_{\mu}\eta_{5}V^{\mu}-mc^{2}]\psi=0,
\end{equation}
where $S_{\mu\nu}=i(\beta_{\mu}\beta_{\nu}-\beta_{\nu}\beta_{\mu})$
is the tensor of
spin, and external potentials are involved in the Lorentz-tensor,
$T^{\mu\nu\lambda}$, and Lorentz-vector, $V^{\mu}$, interactions in
spinor space. Then the equation (\ref{eq13}) is immediately restored
with setting
\begin{equation}\label{eq24}
V^{0}=T^{000}=T^{ij0}=T^{\mu\nu i}=0, \quad
V^{i}=T^{0i0}=-T^{i00}=m\omega x^{i}.
\end{equation}

Taking into account (\ref{eq23}), we conclude that the model of the
DKP oscillator with $b_{1}=1,\quad b_{2}=0$ considered in the paper
\cite{b16} realizes the case of the Lorentz-tensor coupling, whereas
the alternative model with $b_{1}=0,\quad b_{2}=1$ proposed here
includes the coupling of the Lorentz-vector type.

A few remarks concerning the lack of the spin-orbit coupling are in
order. There is another way to introduce the harmonic oscillator in
relativistic quantum mechanics. It has been realized within the
framework of the Dirac equation by means of mixing vector and scalar
harmonic potentials with equal magnitude and equal or opposite sign
(for references see Ref. 10), that results in a quadratic equation
for each spinor component. The energy eigenvalues obtained in the
such approach are characterized by the $(2n+l)$ degeneracy and the
lack of contribution from the spin-orbit interaction just as in our
model. In addition, the lack of the spin-orbit coupling occurs also
in the description of a spin-1 particle with the gyromagnetic ratio
$g=1$ moving in the external electromagnetic field \cite{b23}.

\section*{Acknowledgments}

This work was supported in part by a grant N 0103U000539 from the
Ministry of Education and Science of Ukraine which is gratefully
acknowledged.

\end{document}